\documentclass[%
 aip,
 amsmath,amssymb,
reprint, twocolumn 
]{revtex4-1}

\usepackage{graphicx}
\usepackage{dcolumn}
\usepackage{bm}

\usepackage[utf8]{inputenc}
\usepackage[T1]{fontenc}
\usepackage{mathptmx}
\usepackage{etoolbox}
\usepackage{xcolor}

\newcommand{\fspVec}{{\bm F}\!_{_\text{sp}}}
\newcommand{\fsp}{F\!_{_\text{sp}}}
\newcommand{\fmax}{F\!_{_\text{max}}}

\newcommand{\funit}{k\!_{_B}\!T\!{/}\sigma}

\makeatletter
\def\@email#1#2{%
 \endgroup
 \patchcmd{\titleblock@produce}
  {\frontmatter@RRAPformat}
  {\frontmatter@RRAPformat{\produce@RRAP{*#1\href{mailto:#2}{#2}}}\frontmatter@RRAPformat}
  {}{}
}%
\makeatother
\begin{document}

\preprint{AIP/123-QED}

\title[FI Forces in Active Fluids]{Confinement-Induced Optimization of Fluctuation-Induced Forces in Active Fluids}

\author{Reza Shaebani}
\affiliation{Department of Theoretical Physics and Center for Biophysics, Saarland University, 66123 Saarbr\"ucken, Germany.}

\author{Hashem Fatemi}
\affiliation{School of Quantum Physics and Matter, Institute for Research in Fundamental Sciences (IPM), Tehran 19538-33511, Iran}

\author{Hamidreza Khalilian}
\affiliation{School of Quantum Physics and Matter, Institute for Research in Fundamental Sciences (IPM), Tehran 19538-33511, Iran}

\author{Jalal Sarabadani*}
\affiliation{School of Quantum Physics and Matter, Institute for Research in Fundamental Sciences (IPM), Tehran 19538-33511, Iran}
\email{jalal@ipm.ir}

\date{\today}

\begin{abstract}
Active matter generates nonequilibrium fluctuations that mediate effective interactions between immersed objects. While fluctuation-induced (FI) forces in active fluids depend on activity, density, and geometry, their dependence on confinement remains poorly understood. We study FI forces between fixed intruders in two-dimensional active fluids composed of
self-propelled circular or rodlike particles using Langevin dynamics simulations. We find that the FI force exhibits a pronounced nonmonotonic dependence on intruder separation, reaching a maximum at an optimal gap size well beyond the depletion regime, in contrast to the commonly assumed monotonic decay. This optimal confinement is robust across parameters
and is more pronounced for elongated particles. The effect arises from a confinement-controlled balance between particle transport and crowding: narrow gaps hinder exchange between inner and outer regions, whereas large separations effectively decouple the intruders. At intermediate distances, enhanced crowding around the intruders generates maximal collision-rate asymmetries, leading to the strongest effective interactions. These results identify confinement geometry as a key control parameter for FI forces in active matter.
\end{abstract}

\maketitle

\section{Introduction}

Active matter systems exhibit a wide range of emergent collective phenomena arising from the continuous injection of energy at the microscopic scale  \cite{Cates15,Vicsek12,Shaebani20,Elgeti15}. Over the past two decades, substantial progress has been made in understanding pattern formation, including clustering, segregation, motility-induced phase separation, and self-organization in active-passive mixtures \cite{Stenhammar15,SchwarzLinek12,Gokhale22,Cates15}. In addition, active fluids confined by boundaries or embedded obstacles have been shown to generate effective interactions that can strongly influence structure formation and transport properties  \cite{Paul22,Dor22,Rupprecht18,Makhija16}.

A central class of nonequilibrium interactions in these systems is that of fluctuation-induced (FI) forces, which arise from activity-driven fluctuations in particle density, momentum transfer, and boundary-mediated correlations. Unlike equilibrium Casimir-like forces, which are typically weak and thermally driven, nonequilibrium FI forces can be
significantly enhanced (e.g.\ through self-propulsion and persistence in active matter systems), and can become comparable in magnitude to other mechanical interactions in the system \cite{Angelani11,Liu20,Baek18,Fatemi25,ParraRojas14}. FI forces depend sensitively on system parameters such as particle density, propulsion strength, noise, and the geometry and separation of immersed objects \cite{Harder14,Shaebani13,Cattuto06,Leite16,Ni15,Ray14}.
In particular, effective forces between intruders or confining boundaries have been shown to exhibit transitions between attraction and repulsion as system parameters are varied, highlighting their strongly nonequilibrium nature \cite{Shaebani12,Cattuto06,Ray14,Leite16,Ni15,Feng21}.

Most studies have primarily focused on how FI forces in active systems vary with control parameters such as activity, density, or particle shape. In particular, it is generally assumed that increasing activity enhances the magnitude of effective interactions, while increasing separation between objects leads to a monotonic decay of the force, as the natural generalization of the FI force behavior in passive baths. Indeed, confinement is typically regarded as a passive geometric constraint that sets the scale of interactions, rather than an influential control parameter that can optimize them. However, in a broader class of nonequilibrium transport processes, confinement is known to play a nontrivial and often optimal role. For instance, first-passage times of active agents can exhibit minima
at intermediate confinement lengths \cite{Najafi18,Shaebani20b,Shaebani22}. Such results suggest that geometric constraints may not only restrict dynamics but can also enhance transport and interaction rates under suitable conditions. Whether analogous optimization principles govern FI interactions in active fluids remains largely unexplored.

Here, we address this question by investigating effective forces between two fixed intruders immersed in a two-dimensional active fluid composed of self-propelled circular or rodlike particles. Using extensive Langevin dynamics simulations, we demonstrate that confinement geometry (quantified by the separation between intruders) can act as a powerful tuning
parameter for FI interactions. We show that, in contrast to the commonly assumed monotonic decay with distance, FI forces in active fluids exhibit a pronounced nonmonotonic dependence on intruder separation. Specifically, the force attains a maximum at an intermediate gap size, well beyond the short-range depletion regime.

We find that the optimal confinement is robust across particle shapes and system parameters, although its strength is modulated by activity and particle elongation. This behavior is inherently collective and cannot be explained from dilute or single-particle considerations. It arises from a competition between geometry-induced crowding around the intruders and confinement-controlled particle transport through the gap region, which determines the balance of collision rates on the different sides of the intruders. As a result, the FI force is maximized at intermediate separations, while it is suppressed at both very small and very large separations. Our findings suggest that nonequilibrium fluctuations can generate emergent optimal interaction scales absent in equilibrium systems.

\begin{figure}
\centering
\includegraphics[width=0.99\linewidth]{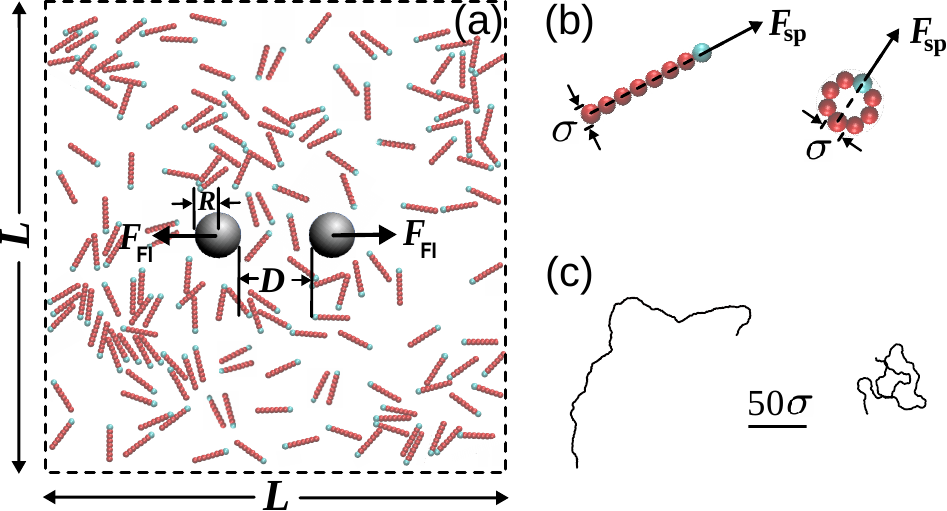}
\caption{(a) Schematic of the two-dimensional active bath composed of self-propelled
circular or rodlike particles interacting through excluded-volume repulsion. Two
immobile circular intruders of radius $R$ are immersed in the bath and separated
by a surface-to-surface distance $D$. The FI force $F_{\mathrm{FI}}$ acting on the intruders is measured along the line connecting their centers in the nonequilibrium steady state. For simplicity, $F_{\mathrm{FI}}$ is denoted by $F$ throughout the remainder of the paper. (b) Each active particle is a rigid composite of touching beads driven by a self-propulsion force $\fspVec$ acting along its intrinsic orientation. Rodlike particles are formed by linear bead chains, whereas circular particles consist of closed bead rings. (c) Representative trajectories of rodlike (left) and circular (right) active particles are shown  over the same time interval for $\fsp\,{=}\,20\,\funit$.}
\label{Fig1}
\end{figure}

\section{Model}

We consider a two-dimensional active fluid composed of self-propelled rigid composite particles confined in a square simulation box of size $L \times L$ with periodic boundary conditions (Fig.\,\ref{Fig1}). Two immobile circular intruders of radius $R$ are embedded in the system and separated by a center-to-center distance $D$. The active bath consists of either rodlike or circular particles constructed from rigidly connected spherical beads of diameter $\sigma$. 
The rod-like partciles have a length of 8$\sigma$, whereas the circular particles have a radius of approximately 1.3$\sigma$, as commonly used in coarse-grained models of active matter \cite{Cates15,Elgeti15}. Because the two geometries occupy different areas, different numbers of composites are employed to achieve the same area fraction.

Each active particle is driven by a constant self-propulsion force $F_{\mathrm{sp}}$ acting along its intrinsic orientation, which defines the direction of propulsion. This converts the system from a passive Brownian fluid into an active nonequilibrium bath with tunable activity. All interactions between beads and between beads and intruders are modeled via a truncated and shifted Lennard-Jones (Weeks-Chandler-Andersen) potential \cite{Weeks71}, ensuring excluded-volume repulsion. The interaction potential between two particles at center-to-center distance $d$ is given by 
\begin{equation}
U_{\mathrm{WCA}}(d) =
\begin{cases}
4\epsilon \left[ \left(\frac{\sigma}{d-\Delta}\right)^{12}
- \left(\frac{\sigma}{d-\Delta}\right)^6 \right]
- U_{\mathrm{LJ}}(d_c), & \hspace{-0.2cm} d-\Delta < d_c, \\
0, & \hspace{-0.2cm} d-\Delta \ge d_c,
\end{cases}
\end{equation}
where $\epsilon$ is the potential well depth, $U_{\mathrm{LJ}}$ is the Lennard-Jones potential, $d_c = 2^{1/6}\sigma$ is the cutoff distance, and $\Delta$ accounts for bead-bead ($\Delta=0$) or bead-intruder ($\Delta = R - \sigma/2$) interactions.

The dynamics of the $i$th bead obeys the Langevin equation
\begin{equation}
M \ddot{\mathbf{r}}_i =
- \eta \dot{\mathbf{r}}_i
- \nabla U_i
+ \boldsymbol{\xi}_i
+ \mathbf{F}^{\mathrm{sp}}_i,
\end{equation}
where $M$ is the bead mass, $\eta$ is the friction coefficient, and $U_i$ is the total interaction potential acting on bead $i$. The stochastic force $\boldsymbol{\xi}_i$ is Gaussian white noise with zero mean and correlations
\begin{equation}
\langle \xi_{i,a}(t)\,\xi_{j,b}(t') \rangle =
4 \eta k_B T\, \delta_{ij}\delta_{ab}\delta(t-t'),
\end{equation}
where $a$ and $b$ denote Cartesian components. Rigid-body constraints are enforced such that the self-propulsion force acts only on a single bead per composite particle (the head bead in rodlike particles or a designated bead in circular particles), ensuring coherent propulsion of each object. Typical particle trajectories shown in Fig.\,\ref{Fig1} indicate a larger effective self-propulsion for rodlike particles compared to circular ones at the same given $F_{\mathrm{sp}}$, since increasing of the effective self-propulsion is associated with enhanced asymptotic diffusion coefficient and, thus, spreading of non-interacting active particles \cite{Shaebani22b,Nossal74,Sadjadi21}.

The bead mass M, diameter $\sigma$, and interaction energy $\epsilon$ are adopted as the fundamental units of mass, length, and energy, respectively. The solvent friction coefficient is fixed at $\eta = 10$ throughout all simulations. Langevin dynamics simulations are performed using the LAMMPS package \cite{LAMMPS,Plimpton95}. The equations of motion are integrated with a time step of $10^{-4}\tau$, where $\tau = \sqrt{M\sigma^{2}/\epsilon}$ defines the characteristic simulation time scale.

In the steady state, that is reached after $10^7$ time steps in the presence of active forces, we measure the net force exerted by the active bath on each intruder along the line connecting their centers. This FI force is denoted by $F$ in the following. Due to strong temporal fluctuations, forces are obtained by long-time averaging over $10^4$ steady-state successive time intervals and further ensemble averaging over $10^2$ independent realizations.

\section{Results and Discussion}

\subsection{Nonmonotonic fluctuation-induced forces under confinement}

\begin{figure}
\centering
\includegraphics[width=0.85\linewidth]{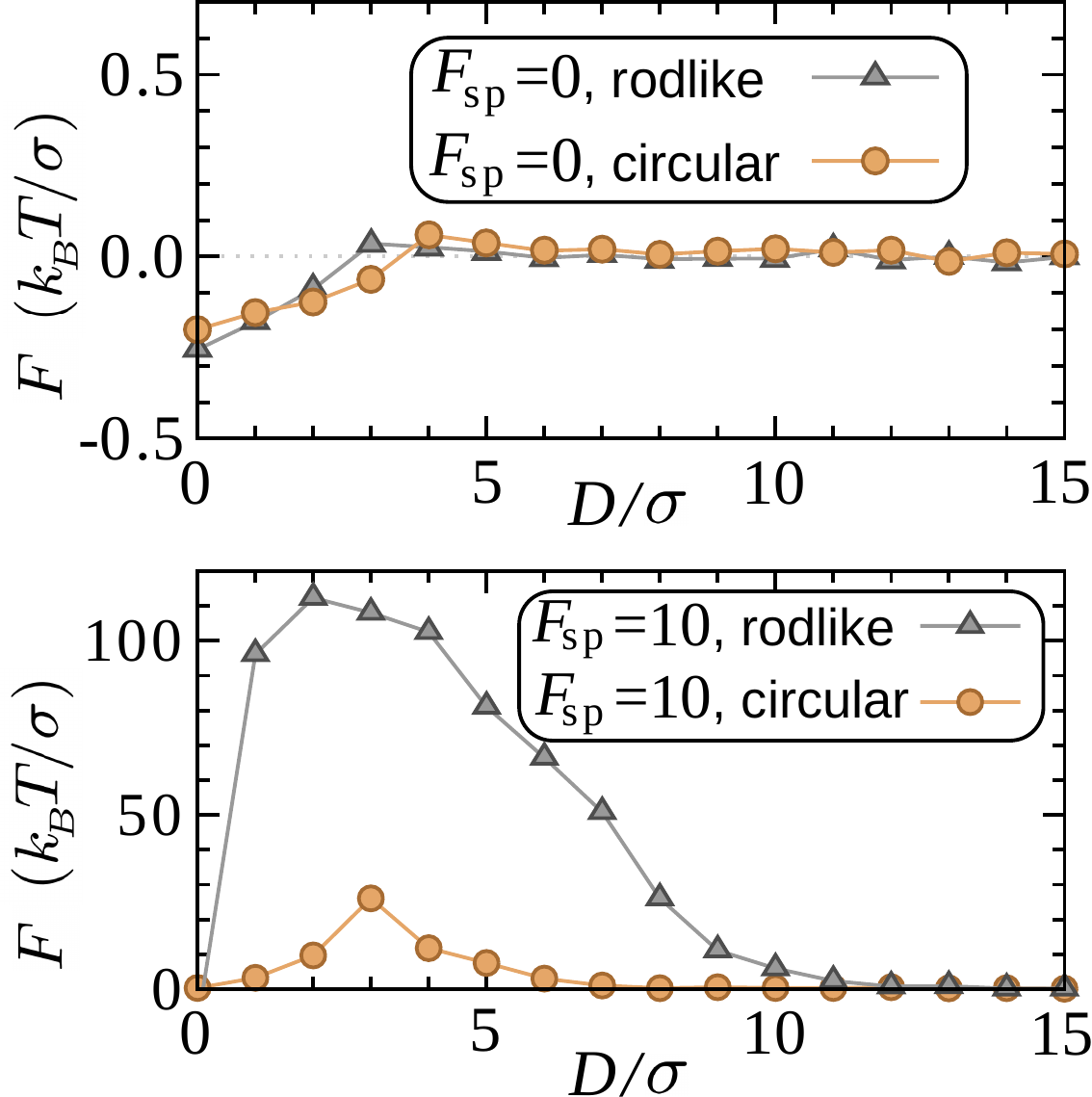}
\caption{FI force $F$ between the intruders as a function of the scaled
gap size $D\,{/}\,\sigma$ for baths of passive ($\fsp{=}\,0$; upper panel) and active
($\fsp{=}\,10\,\funit$; lower panel) circular or rodlike particles. The area fraction
is $\phi\,{=}\,0.1$. Error bars are smaller than the symbol size. In contrast to the
passive case, active particles induce a pronounced nonmonotonic dependence of the
force on the intruder separation.}
\label{Fig2}
\end{figure}

\begin{figure}
\centering
\includegraphics[width=0.99\linewidth]{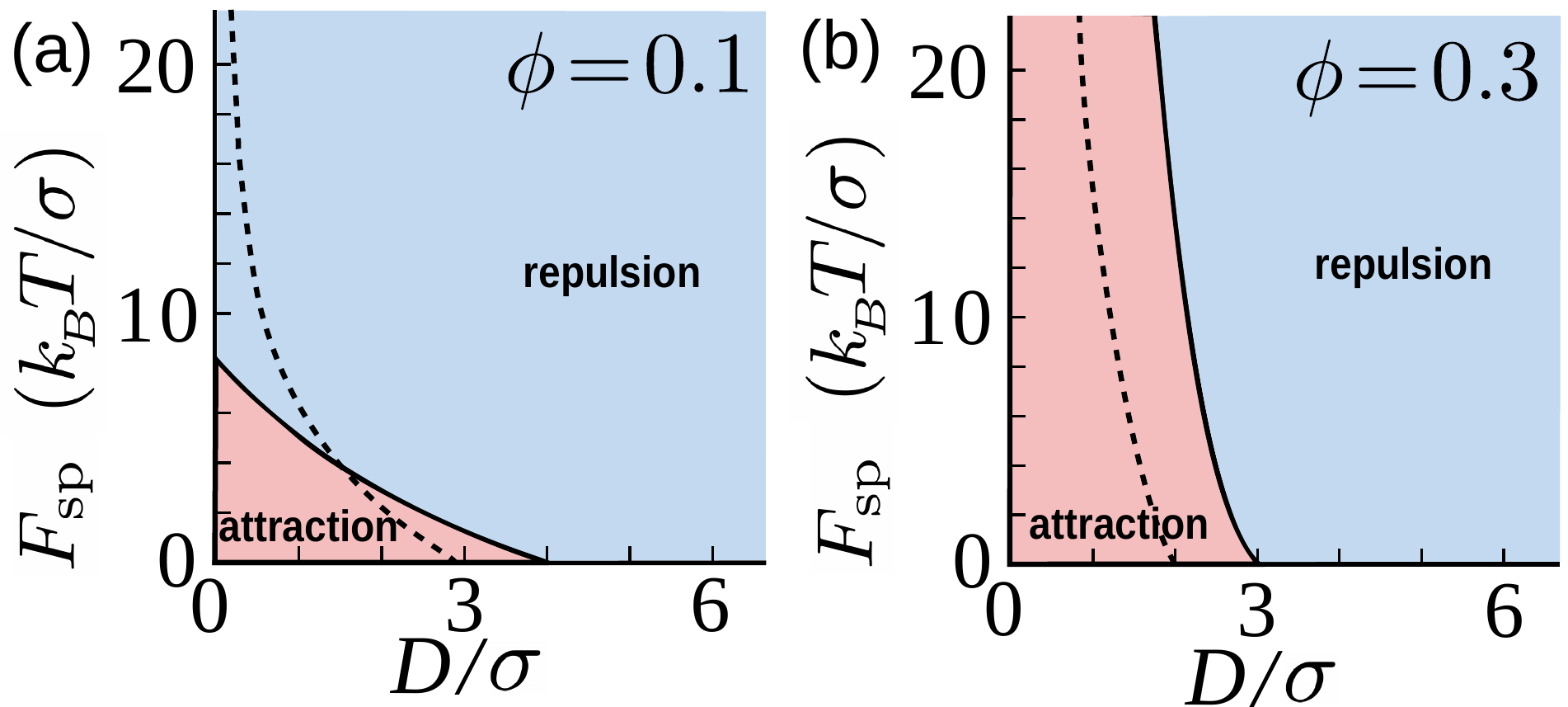}
\caption{Phase diagram of FI force sign (attraction vs repulsion) in the
$(D,\fsp)$ plane for bath densities (a) $\phi\,{=}\,0.1$ and (b) $\phi=0.3$.
The solid (dashed) line marks the transition between attractive and repulsive regimes
for circular (rodlike) active particles.}
\label{Fig3}
\end{figure}

We first compare the behavior of Casimir-like interactions emerged between immobile intruders in passive and active baths. Figure~\ref{Fig2} shows the FI force $F$ between the two intruders as a function of their separation $D$ for passive and active baths composed of either circular or rodlike particles.

For passive systems ($F_{\mathrm{sp}}{=}0$), the FI force is relatively weak, with a magnitude below $1\,\funit$ for the parameters considered here. As shown in Fig.\,\ref{Fig2} (upper panel), the force changes sign from attraction to repulsion upon increasing the separation between the intruders. This behavior is consistent with previous studies of FI forces in confined passive systems driven by stochastic fluctuations \cite{Shaebani12,Shaebani13,Cattuto06}. The interaction remains short-ranged and rapidly approaches zero beyond the depletion regime.

The situation changes qualitatively in active fluids. An example is shown in the lower panel of Fig.\,\ref{Fig2} for $F_{\mathrm{sp}}=10\,\funit$. In this case, the FI force becomes orders of magnitude larger than in the passive bath and exhibits a pronounced nonmonotonic dependence on the intruder separation. Rather than decaying monotonically with distance, the force first increases, reaches a maximum value $F_{\max}$ at an intermediate separation, and only then decreases at larger distances. The position of the maximum lies well beyond the short-range depletion interaction range, demonstrating that the observed effect originates from nonequilibrium active fluctuations rather than simple excluded-volume mechanisms.

A comparison between circular and rodlike active particles reveals important shape-dependent differences. Under otherwise identical conditions, rodlike particles generate substantially stronger FI forces than circular particles. For example, at $F_{\mathrm{sp}}=10\,\funit$, the maximum force in rodlike baths exceeds $100\,\funit$, whereas considerably smaller values are obtained for circular particles. There is also a slight difference between the optimal distance for circular and rodlike particles. The stronger interaction generated by rods reflects the enhanced persistence and steric correlations associated with particle  elongation. Nevertheless, both particle shapes display the same qualitative feature: the existence of an optimal intruder separation that maximizes the effective interaction. The emergence of an optimal separation is the central finding of this work. The results indicate that confinement is not merely a geometric parameter setting the interaction range, but rather an active control parameter that can enhance the strength of FI forces.

The sign of the FI force is determined by the interplay between activity, bath density, particle shape, and intruder separation. Figure\,\ref{Fig3} summarizes the force sign switching behavior in the $(D,F_{\mathrm{sp}})$ plane for two representative bath densities. For sufficiently weak activity, the force changes from attraction to repulsion as the separation increases, similar to the behavior of FI interactions in passive systems \cite{Shaebani12,Cattuto06}. Increasing the propulsion strength shifts the crossover separation to smaller values and eventually eliminates the attractive regime, yielding purely repulsive interactions. This trend is observed for both circular and rodlike particles,
although the phase boundaries differ quantitatively.

\begin{figure}
\centering
\includegraphics[width=0.8\linewidth]{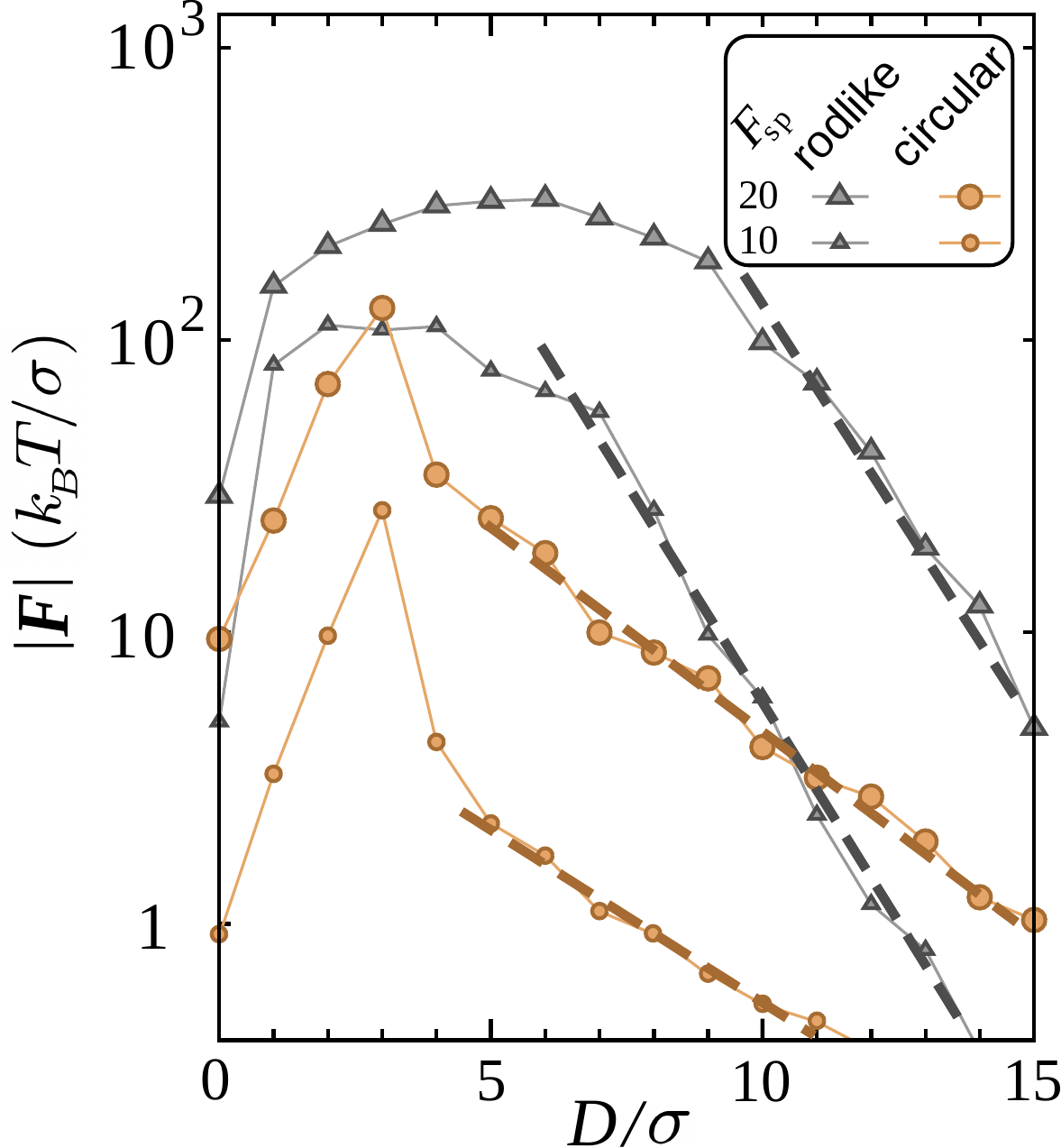}
\caption{FI force $F$ versus the scaled gap size $D\,{/}\,\sigma$ for
circular and rodlike active particles at different self-propulsion forces $\fsp$. The
area fraction is $\phi\,{=}\,0.1$. The dashed lines show exponential fits to the
large-distance tail, $F_{\mathrm{tail}}\,{\sim}\,\exp(-\alpha \,D\,{/}\,\sigma)$.}
\label{Fig4}
\end{figure}

The bath density also influences the force sign reversal. At the higher density $\phi=0.3$, the transition to purely repulsive forces occurs at larger propulsion strengths than for $\phi=0.1$. Thus, increasing density tends to stabilize the attractive regime. This behavior is consistent with the stronger confinement-induced crowding that develops in denser active baths. Figure\,\ref{Fig3} further shows that the dependence of the crossover boundary on activity is more pronounced for circular particles than for rodlike ones.

To characterize the range of the interaction, Fig.\,\ref{Fig4} presents $F$ as a function of separation for different values of the propulsion force. The large-distance tails can be reasonably described by an exponential form 
\begin{equation}
F_{\mathrm{tail}} \sim \exp(-\alpha \, D/\sigma),
\end{equation}
where $\alpha$ is an effective decay exponent. Increasing activity strongly amplifies the force over the entire separation range and slightly shifts the location of the maximum toward larger distances, particularly for rodlike particles. However, the overall shape of the curves remains qualitatively unchanged, indicating that the mechanism responsible for force optimization is robust.

\begin{figure}[t]
\centering
\includegraphics[width=0.85\linewidth]{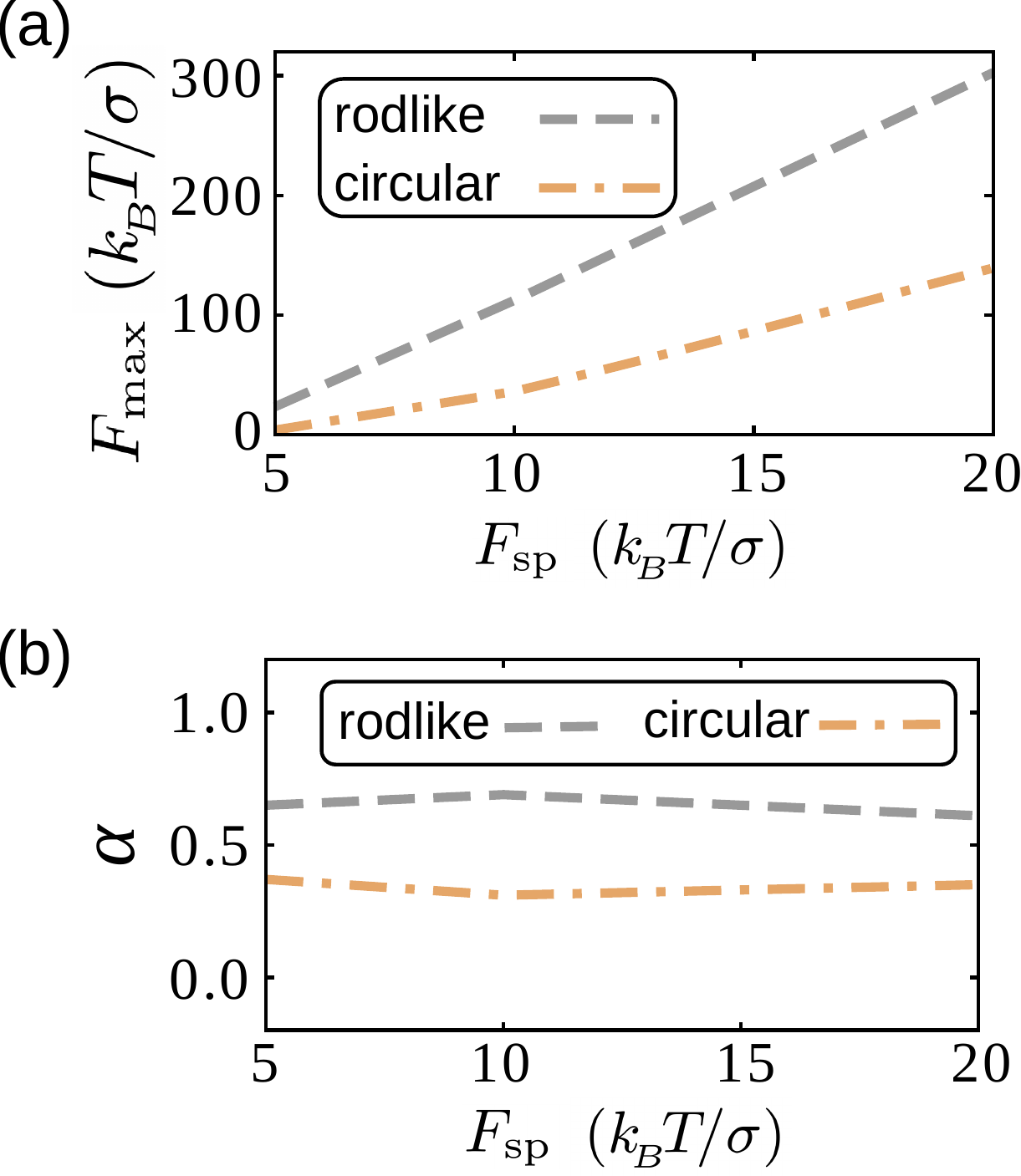}
\caption{(a) Maximum FI force $\fmax$ and (b) decay exponent $\alpha$ versus the self-propulsion force $\fsp$ for circular and rodlike activeparticles at $\phi=0.1$. The exponent $\alpha$ characterizes the exponential decay of the force at large intruder separations.}
\label{Fig5}
\end{figure}

The dependence of the peak force $F_{\max}$ on activity is summarized in the upper panel of Fig.\,\ref{Fig5}. For both particle shapes, $F_{\max}$ increases approximately linearly with the propulsion force. The increase is considerably stronger for rodlike particles, leading
to an ever-growing difference between rodlike and circular baths as activity increases. This nearly linear scaling implies that the magnitude of FI forces can become arbitrarily large in sufficiently active systems and may therefore compete with or even dominate other interactions relevant for self-organization and assembly.

The lower panel of Fig.\,\ref{Fig5} shows the corresponding decay exponent $\alpha$. In contrast to the strong variation of $F_{\max}$, the interaction range exhibits only a weak dependence on activity. The values of $\alpha$ remain approximately constant over the investigated propulsion range, indicating that activity primarily controls the amplitude rather than the characteristic decay length of the force. We also find that $\alpha$ is systematically smaller for circular particles than for rodlike particles, corresponding to weaker but longer-ranged interactions. Rodlike particles therefore generate stronger forces, whereas circular particles mediate interactions over larger distances.

\begin{figure*}[htb]
\centering
\includegraphics[width=0.99\linewidth]{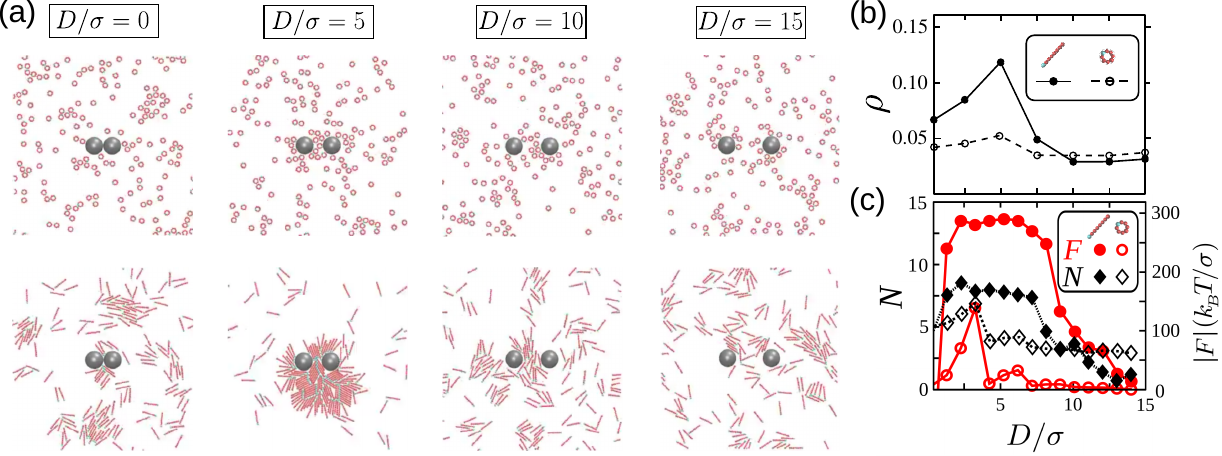}
\caption{Microscopic origin of the confinement-dependent FI force. (a) Typical steady-state particle configurations around the intruders for circular and rodlike active baths at several gap sizes and $\phi\,{=}\,0.1$ and $\fsp{=}\,10\,\funit$. (b) Mean number density of active particles in the vicinity of the intruders as a function of intruder separation $D$. (c) FI force $F$ and corresponding average collision number $N$ versus separation $D$.}
\label{Fig6}
\end{figure*}

The results presented in Figs.\,\ref{Fig2}-\ref{Fig5} demonstrate that active fluids generate long-range FI forces whose magnitude, sign, and range can be tuned through activity, density, particle shape, and confinement. Most importantly, they reveal the existence of an optimal intruder separation at which the force reaches a maximum. In the following subsection, we investigate the microscopic origin of this nonmonotonic distance dependence by analyzing particle density distributions and collision statistics around the intruders.

\subsection{Collision statistics and microscopic origin of the confinement-induced force maximum}

To uncover the mechanism responsible for the nonmonotonic dependence of the FI force on intruder separation, we analyze the spatial organization of active particles around the intruders together with the corresponding collision statistics. In passive nonequilibrium systems, effective long-range interactions have been linked to FI modifications of local pressure fields imposed by confining boundaries \cite{Shaebani12,Cattuto06}. In active fluids, forces acting on immersed objects have similarly been associated with asymmetric particle accumulation and collision-induced momentum transfer near confining surfaces \cite{Paul22}. Motivated by these observations, we investigate whether the force maximum observed here can be traced back to a confinement-dependent redistribution of active particles around the intruders.

Figure~\ref{Fig6}(a) shows typical snapshots of particle configurations around the intruders for circular and rodlike active baths at different gap sizes. It shows that active particles expectedly tend to accumulate around the intruders. This can lead to the formation of highly dynamic clusters around each intruder. These structures continuously form, reorganize, and dissolve due to the persistent propulsion of the bath particles. As a result, the intruders are surrounded by fluctuating particle crowds that mediate momentum transfer from the active fluid. In such an environment, the intruder separation interestingly plays a nontrivial 
role:\ The extent of particle accumulation depends strongly on the available gap between the intruders. When the separation is very small, the confined region cannot efficiently accommodate active particles and only weak crowding develops. Increasing the separation facilitates particle penetration into the gap and promotes the formation of dense accumulations around both intruders. The stochastic interaction between the crowds around both intruders can destabilize and disrupt the crowds, enhancing the particle-intruder collisions. At even larger separations, however, the influence of one intruder on the particle distribution around the other gradually decreases and the two crowds become effectively independent. Consequently, the local structure approaches that of an isolated intruder. Figure~\ref{Fig6}(b) represents the average particle density in the vicinity of the intruders as a function of their separation. It confirms that for both circular and rodlike active particles, the local density exhibits a clear nonmonotonic dependence on $D$, reaching a maximum at intermediate separations. The effect is more pronounced for rodlike particles. 

To establish a direct connection between particle accumulation and the effective force, we measure the average number of particle-intruder collisions as well as the effective force. As shown in Fig.\,\ref{Fig6}(c), the collision number $N$ closely follows the same nonmonotonic trend as the FI force. Both quantities increase from small separations, attain a maximum at intermediate $D$, and subsequently decrease when the intruders are moved further apart. The strong correspondence between $N$ and $F$ indicates that the force is governed primarily by collision-mediated momentum transfer from the active bath. The observed force maximum therefore originates from a competition between two opposing confinement effects. For narrow gaps, particle transport between the inner and outer regions is strongly hindered, suppressing the formation of dense active crowds and limiting collision events. For very large separations, the intruders become effectively decoupled and the correlations induced by the active bath weaken. Between these two limits, confinement optimally balances 
particle accumulation and transport, maximizing collision asymmetries and giving rise to the strongest FI interaction.

Particle shape further modulates this mechanism. Rodlike particles produce larger density enhancements and higher collision rates than circular particles at all separations considered. Their elongated geometry promotes local alignment and increases the residence time of particles near the intruder surfaces, leading to more persistent accumulations and stronger momentum transfer. This explains why rodlike active baths generate significantly larger FI forces throughout the parameter range studied.

\section{Conclusions}

We have investigated FI forces between two immobile intruders immersed in two-dimensional active fluids composed of self-propelled circular or rodlike particles. Using Langevin dynamics simulations, we demonstrated that the effective interaction exhibits a pronounced nonmonotonic dependence on the separation between the intruders. Rather than decreasing monotonically with distance, the force reaches a maximum at an intermediate separation that lies well beyond the short-range depletion regime. This behavior is observed over a broad range of activities and bath densities and represents a robust feature of active FI interactions.

The magnitude of the force depends on both activity and particle shape. Rodlike particles 
generate substantially stronger forces than circular particles, while circular particles 
generally produce interactions with a longer spatial range. In addition, we identified 
the conditions under which the force changes sign between attraction and repulsion and 
mapped the corresponding regions in the parameter space of activity and intruder separation.

To elucidate the physical origin of the force maximum, we analyzed density distributions and collision statistics around the intruders and verified that the FI force is directly correlated with the average number of particle-intruder collisions. The observed optimum results from a competition between crowd formation and particle transport through the confined region between the intruders. 

Our results identify confinement geometry as a key control parameter for FI interactions in active matter. More generally, they demonstrate that nonequilibrium active fluctuations can generate emergent optimal interaction scales that have no equilibrium counterpart. These findings provide new insight into the design of active materials, confined active fluids, and self-assembled structures whose effective interactions can be tuned through geometry alone.

\begin{acknowledgments}
This work is dedicated to the memory of Rudolf Podgornik, whose enduring inspiration, generous mentorship, and profound scientific insight greatly influenced J.S.'s research journey and understanding of soft condensed matter and statistical physics.

R.S.\ acknowledges support by the Deutsche Forschungsgemeinschaft (DFG) through Collaborative Research Center SFB 1027 and by the Young Investigator Grant of Saarland University, Grant No.\ 7410110401. 
\end{acknowledgments}

\section*{Competing interests}

The authors declare that they have no competing interests.

\section*{Data Availability Statement}

The data that support the findings of this study are available within the article.

\bibliography{Refs-ActiveCasimir-CirularMotion}

\end{document}